\documentstyle{prhep97}

\newcommand{\mt}{m_{\tau}}

\makeatletter
\let\chapter\hid@chapter
\makeatother
\begin{document}

%\pagenumbering{empty}

% The following definitions need to be customised;

% Will appear on page headings
\authorrunning{J.H.~K\"uhn and E.~Mirkes}
\titlerunning{{\talknumber}: CP Violation in  Tau Decays}
 
% Now the full name of author and talk

% For plenary talks, the talk number is that of the session
\def\talknumber{903} 
\title{{\talknumber}:
       CP Violation from Charged Higgs Exchange in Hadronic Tau Decays
       with Unpolarized Beams}
\author{J.H.~K\"uhn and \underline{E.~Mirkes}}
\institute{Institut f\"ur Theoretische Teilchenphysik, 
         Universit\"at Karlsruhe,\\ D-76128 Karlsruhe, Germany}
\maketitle
\vspace{-6cm}
\hfill \vtop{   \hbox{\bf hep-ph/9711364}
                \hbox{\bf TTP97-48}}\footnote{Talk given 
by E. Mirkes at the International Europhysics Conference on High-Energy Physics
(HEP 97), Jerusalem, Israel, 19-26 Aug 1997. }
\vspace{6cm}
\begin{abstract}
CP violating signals in semileptonic $\tau$ decays induced
by an exotic scalar exchange are studied in a completely
model-independent way.
These can be observed in decays of unpolarized single $\tau$'s
even if their rest frame cannot be reconstructed. No beam polarization 
is required. The importance of the two-meson channel, in particular the 
$K\pi$ final state is emphasized.
\end{abstract}
%
%%%%%%%%%%%%%%%%%%%%%%%%%%%%%%%%%%%%%%%%%%%%%%%%%%%%%%
\section{Introduction}
%%%%%%%%%%%%%%%%%%%%%%%%%%%%%%%%%%%%%%%%%%%%%%%%%%%%%%
CP violation has been experimentally observed only in the
$K$ meson system. The effect can be explained by a nontrivial complex
phase in the CKM flavour mixing matrix 
\cite{kmmat}. However, the fundamental origin of this CP violation
is still unknown.
In particular the CP properties of the third fermion
family are largely unexplored.
Production and decay of $\tau$ leptons might offer a particularly clean
laboratory to study these effects.
In this contribution, we investigate the effects of
CP violation  \cite{cp1} to be observed in semileptonic $\tau$ decays
which could arise in a framework outside the mechanism prosposed
by Kobayashi and Maskawa.
We show that the structure function
formalism of Ref.~\cite{km1} allows for a systematic analysis
of possible CP violation effects in the two and three
meson channels. Special emphasis is put on the $\Delta S=1$ transition
$\tau\rightarrow K\pi\nu_\tau$ where possible CP violating signals from
multi Higgs boson models \cite{mhiggs}  would be signaled by a nonvanishing
difference between the structure functions
$W_{SF}[\tau^-\rightarrow (K\pi)^-\nu_\tau]$ and
$W_{SF}[\tau^+\rightarrow (K\pi)^+\nu_\tau]$.
Such a measurement is possible for unpolarized single $\tau$'s without
reconstruction of the $\tau$ rest frame and without
polarized incident $e^+e^-$ beams.
It is shown that this CP violation requires both
nonvanishing hadronic phases and CP violating phases in the
Hamiltonian,
where the hadronic phases arise from the interference of complex 
Breit-Wigner propagators, whereas the CP violating phases could 
arise from an exotic charged Higgs boson.
An additional independent test of CP violation
in the two meson case is possible, but
would require the knowledge of the full kinematics and  $\tau$
polarization. 

%%%%%%%%%%%%%%%%%%%%%%%%%%%%%%%%%%%%%%%%%%%%%%%%%%%%%%
\section{CP Violating Signals in the $\tau\rightarrow K\pi\nu_\tau$ Decay Mode}
%%%%%%%%%%%%%%%%%%%%%%%%%%%%%%%%%%%%%%%%%%%%%%%%%%%%%%
Transitions from the vacuum to two pseudoscalar mesons $h_1$ and
$h_2$ 
are induced through vector and scalar currents only.
%  , where the latter can 
%  be related to the former with the help of Eq.~(\ref{motion}).
Expanding these hadronic matrix elements along the set of independent momenta
$(q_1-q_2)_\beta$ and $Q_\beta=(q_1+q_2)_\beta$ we define
$(T_{\alpha\beta}=  g_{\alpha \beta} - (Q_\alpha Q_\beta)/Q^2)$
\begin{eqnarray}
%J_\beta=\langle h_1(q_1)h_2(q_2)|\bar{u}\gamma_\beta d|0 \rangle=
\langle h_1(q_1)h_2(q_2)|\bar{u}\gamma_\beta d|0 \rangle&=&
(q_1-q_2)^\alpha\,T_{\alpha\beta}\,\,F(Q^2) \,+\,
Q_\beta\, F_S(Q^2)\\
\langle h_1(q_1)h_2(q_2)|\bar{u} d|0 \rangle &=& F_H(Q^2).
\label{jdef}
\end{eqnarray}
The representation of the hadronic amplitude
$\langle h_1h_2|\bar{u}\gamma_\beta d|0\rangle $
corresponds to a decomposition into spin one and spin
zero contributions,
{\it e.g.} the vector form factor $F(Q^2)$
corresponds to the $J^P = 1^-$ component
of the  weak charged current, and the scalar form factor 
$F_S(Q^2)$ to the $J^P = 0^+$ component. 
A fictitious scalar Higgs exchange contribution is proportional to $\eta_S F_H$.
The general amplitude for the $\Delta S=1$ decay (where 
$h_1\equiv K, h_2\equiv \pi$)
\begin{equation}
\tau^-(l,s) \rightarrow \nu(l^\prime,s^\prime) + h_1(q_1,m_1)+h_2(q_2,m_2)\>,
\label{twoh}
\end{equation}
can thus be written as
\begin{equation}
{\cal{M}}=
\sin\theta_{c}\frac{\,G}{\sqrt{2}}
  \bar{u}(l^\prime,s^\prime)\,\gamma_{\alpha}
           (1-\gamma_5)\,u(l,s) 
  \left[(q_1-q_2)_\beta\,T^{\alpha\beta}\, F   \, + \,
  Q^\alpha\,\tilde{F}_S\right]
\label{me2h}
\end{equation}
\vspace*{-3mm}
\begin{equation}
\mbox{with}\hspace{1cm}\tilde{F}_S = F_S \,+\, \frac{\eta_S}{m_\tau}\,F_H
\label{fs}
\end{equation}
In Eq.~(\ref{me2h}) $s$ 
denotes the polarization 4-vector of the $\tau$ lepton.
The complex parameter $\eta_S$ in Eq.~(\ref{fs}) 
transforms like 
\begin{eqnarray}
\eta_S & \stackrel{\mbox{CP}}{\longrightarrow}&\eta^\ast_S
\end{eqnarray}
and thus allows for the parametrization of possible CP violation.
Up to the small isospin breaking terms, induced for example
by the small quark mass difference, CVC implies the
vanishing of $F_S$ for the two pion ($h_1\equiv\pi^-,h_2\equiv\pi^0$) case. 
For the
transition $\tau\rightarrow K\pi\nu$ the $J=1$ form factor $F$
is dominated by the $K^*(892)$ vector resonance contribution. 
The scalar form factor $F_S$ is expected to
receive a sizable resonance contribution 
($\sim 5\%$ to the decay rate)
from the $K_0^*(1430)$ with $J^P=0^+$
\cite{kpi}.
The corresponding $\tau^+$ decay is obtained from 
Eq.~(\ref{me2h})
through the substitutions 
\begin{equation}
(1-\gamma_5)\, \rightarrow \, (1+\gamma_5), \hspace{1cm}
%\chi      \, \rightarrow \, \chi^*,     \hspace{1cm}
  \eta_S      \, \rightarrow \, \eta_S^*.     \hspace{1cm}
%\eta_S^d      \, \rightarrow \, \eta_S^{d *}.     \hspace{1cm}
\end{equation}

Reaction (\ref{twoh}) is most easily analyzed in the hadronic rest frame
$\vec{q}_{1}+\vec{q}_{2}=~0$ \cite{km1}.
After integration over the unobserved
neutrino direction,
the differential decay rate in the
rest frame of $h_{1}+h_{2}$ is given   by \cite{kpi,km1}
\begin{eqnarray}
d\Gamma(\tau^-\rightarrow 2h\nu_\tau)&=&
 \left\{
    \bar{L}_{B} {W}_{B} 
%  + \bar{L}_{SA}{W}_{SA}(1+\tilde\eta_S^d)^2
  + \bar{L}_{SA}{W}_{SA}
  + \bar{L}_{SF}{W}_{SF} 
%  + \bar{L}_{SG}{W}_{SG})(1+\tilde\eta_S^d)  \right\}\nonumber\\[2mm]
  + \bar{L}_{SG}{W}_{SG}  \right\}\nonumber\\[3mm]
&&
\hspace{-2cm}
\frac{G^{2}}{2\mt} 
\sin^2\theta_{c}
\frac{1}{(4\pi)^{3}}
\frac{(\mt^{2}-Q^{2})^{2}}{\mt^{2}}\,
\,|\vec{q}_{1}|\,\,
     \frac{dQ^{2}}{\sqrt{Q^{2}}}  \, \frac{d\cos\theta}{2}\,
     \frac{d\alpha}{2\pi}\,
     \frac{d\cos\beta}{2}\label{gamma}\>.
\end{eqnarray}
For the definition and discussion of the angles and leptonic
coefficients $\bar{L}_X$ in Eq.~{(\ref{gamma})}
we refer the reader to Ref.~\cite{cp1}.
The coefficients $\bar{L}_X$ contain all  $\alpha, \beta$ and $\theta$
angular  and $\tau$-polarization dependence.
The hadronic structure functions $W_X$,
$X\in\{B,SA,SF,SG\}$, depend only on
$Q^2$ and the form factors $F$ and $\tilde{F}_S$ of the hadronic current.
One has \cite{kpi,km1,cp1}:
\begin{equation}
\begin{array}{ll}
W_B[\tau^-]    = 4 (\vec{q}_1)^2\,|F|^2 &
\hspace{1cm}W_{SF}[\tau^-]
 = 4\sqrt{Q^2}|\vec{q}_1|\, \mbox{Re}\left[F\tilde{F}_S^*\right]       \\
W_{SA}[\tau^-] = Q^2\,  |\tilde{F}_S|^2 
&
\hspace{1cm}W_{SG}[\tau^-]
 =-4\sqrt{Q^2}|\vec{q}_1|\, \mbox{Im}\left[F\tilde{F}_S^*\right]
\end{array}
\label{wsg}
\end{equation}
The hadronic structure functions $W_X[\tau^+]$  are
obtained by the replacement $\eta_S\rightarrow\eta_S^*$
in $\tilde{F}_S$ in Eqs.~(\ref{wsg},\ref{fs}).
CP conservation implies that all four structure functions are identical
for $\tau^+$ and $\tau^-$. With the ansatz 
for the form factors formulated in Eq.~(\ref{me2h})
CP violation can be present in $W_{SF}$ and $W_{SG}$
only and requires complex $\eta_S$.

As demonstrated in Ref.~\cite{cp1} $W_{SF}$ can be measured in 
$e^+e^-$ annihilation experiments in the study of single unpolarized
$\tau$ decays even if the $\tau$ rest frame cannot be reconstructed.
In this respect the result differ from earlier studies of the two 
meson modes where either polarized beams and reconstruction
of the full kinematics \cite{tsaicp} or correlated fully reconstructed
$\tau^-$ and $\tau^+$ decays were required \cite{nelson1}.
The determination of $W_{SG}$, however, requires the knowledge of the full
$\tau$ kinematics and $\tau$  polarization \cite{cp1}
which is possible with the help of vertex detectors.
The corresponding distributions in this second case are 
equivalent to the correlations proposed in Refs.~\cite{tsaicp,nelson1}.

The crucial observation made in Ref.~\cite{cp1} is that
one can measure the following CP-violating differences
\begin{equation}
\Delta W_{SF} = \frac{1}{2}
\left( W_{SF}[\tau^-] -  W_{SF}[\tau^+]\right)
\hspace{5mm}
\Delta W_{SG} = \frac{1}{2}
\left( W_{SG}[\tau^-] -  W_{SG}[\tau^+]\right)
\end{equation}
under the above mentioned conditions.
The hadronic structure functions
$W_X[\tau^-]$ and $W_X[\tau^+]$ differ only in the phase os
$\eta_S$ and one obtains
\begin{equation}
\Delta W_{SF}
\,\,= \,\,\, 4\sqrt{Q^2}|\vec{q}_1|\, \,
%\frac{Q^2}{m_{\tau}(m_u-m_d)}\,\,\,\mbox{Im}\left(FF_S^*\right)\,\,
\frac{1}{m_{\tau}}\,\,\mbox{Im}\left(F F_H^*\right)\,\,
              \,     \mbox{Im}\left(\,\eta_S \right) \>.
\label{dwsf}
\end{equation}
In essence the measurement on $\Delta W_{SF}$ 
analyses the difference in the
correlated energy distribution
of the mesons $h_1$ and $h_2$ from $\tau^+$ and $\tau^-$
decay in the laboratory. 
As already mentioned, $\Delta W_{SF}$ is observable for single
$\tau^+$ and $\tau^-$ decays without knowledge of the
$\tau$ rest frame. Any nonvanishing experimental result for 
$\Delta W_{SF}$ would be a clear signal of CP violation.
Note that a nonvanishing $\Delta W_{SF}$ requires nontrivial hadronic phases
(in addition to the CP violating phases $\eta_S$) in the form factors
$F$ and $F_H$. Such hadronic phases in $F$ ($F_H$) originate
in the $K\pi\nu_\tau$ decay mode from complex Breit Wigner
propagators for the $K^*$ ($K_0^*$) resonance. Sizable effects
of these hadronic phases are expected in this decay mode \cite{kpi}.

Once the $\tau$ rest frame is known and a preferred direction
of polarization exists
one may also determine $\Delta W_{SG}$ which is
theoretically given by
\begin{equation}
\Delta W_{SG}
\,\,= \,\,\, 4\sqrt{Q^2}|\vec{q}_1|\,\,\,
%\frac{Q^2}{m_{\tau}(m_u-m_d)}
\frac{1}{m_{\tau}}
\,\,\, \mbox{Re}\left(F F_H^*\right)
              \,\,     \mbox{Im}\left(\eta_S \right) \>.
\label{dwsg}
\end{equation}
Any observed nonzero value of $\Delta W_{SF},\Delta W_{SG}$ would signal a 
true CP violation. Eqs.(\ref{dwsf}) and (\ref{dwsg}) show that the sensitivity
to CP violating effects in $\Delta W_{SF}$ and $\Delta W_{SG}$
can be fairly different depending on the hadronic phases.
Whereas $\Delta W_{SF}$ requires nontrivial hadronic phases
$\Delta W_{SG}$ is maximal for fixed $\eta_S$ in the absence of hadronic
phases.
\vspace*{-2mm}

\section{Three  Meson Decays}
\vspace*{-2mm}
%%%%%%%%%%%%%%%%%%%%%%%%%%%%%%%%%%%%%%%%%%%%%%%%%%%%%%
The structure function formalism \cite{km1}
allow also for a systematic analysis of possible CP
violation effects in the three meson case.
Some of these effects have already been briefly discussed in
Ref.~\cite{argonne}. The $K\pi\pi$ and $KK\pi$ decay modes
with nonvanishing vector \underline{and} axial vector
current are of particular importance for the detection of possible
CP violation originating from exotic intermediate vector bosons.
This would be signalled by a nonvanishing difference between
the structure functions $W_X(\tau^-)$ and $W_X(\tau^+)$
with $X\in\{F,G,H,I\}$. A difference in the structure functions
with $X\in\{SB,SC,SD,SE,SF,SG\}$ can again be induced through
a CP violating scalar exchange.
CP violation in the three pion channel has been also discussed in 
Ref.~\cite{hagiwara} and in the $K\pi\pi$ and $KK\pi$ channels in 
Ref.~\cite{koerner},
where the latter analysis is based on the ``$T-$odd'' correlations 
in Ref.~\cite{km1} and the vector meson dominance parametrizations
in Ref.~\cite{roger1}.

\vspace*{-2mm}
% ---- Bibliography ----
%

\end{document}